\documentclass[10pt,letterpaper]{article}
\usepackage[top=0.85in,left=2.75in,footskip=0.75in,marginparwidth=2in]{geometry}
\usepackage[utf8]{inputenc}
\usepackage{nameref,hyperref}
\usepackage[right]{lineno}
\usepackage{xcolor}
\usepackage{microtype}

\DisableLigatures[f]{encoding = *, family = * }
\raggedright
\usepackage{changepage}
\usepackage{comment}
\usepackage[aboveskip=1pt,labelfont=bf,labelsep=period,singlelinecheck=off]{caption}

\makeatletter
\renewcommand{\@biblabel}[1]{\quad#1.}
\makeatother

\usepackage{lastpage,fancyhdr,graphicx}
\usepackage{epstopdf}
\fancyheadoffset[L]{2.25in}
\fancyfootoffset[L]{2.25in}

\usepackage{color}

\definecolor{Gray}{gray}{.25}

\usepackage{graphicx}

\usepackage{sidecap}

\usepackage{wrapfig}
\usepackage[pscoord]{eso-pic}
\usepackage[fulladjust]{marginnote}
\reversemarginpar
\usepackage[T1]{fontenc}
\usepackage{amssymb}
\usepackage{amsmath}
\usepackage[font=small,labelfont=bf]{caption} 
\usepackage[format=hang,singlelinecheck=0,font={sf,small},labelfont=bf,caption=false]{subfig}
\usepackage[authoryear]{natbib}

\makeatletter
\newcommand{\manuallabel}[2]{\def\@currentlabel{#2}\label{#1}}
\makeatother

\begin{document}
\vspace*{0.35in}
\manuallabel{fig:simulation_and_bootstrap}{S1}

\begin{flushleft}
{\Large
\textbf{Fundamental Law of Memory Recall}
}
\newline
\\
Michelangelo Naim\textsuperscript{1+},
Mikhail Katkov\textsuperscript{1+},
Sandro Romani\textsuperscript{2},
Misha Tsodyks\textsuperscript{1,3*}
\\
\bigskip
\bf{1} Department of Neurobiology, Weizmann Institute of Science, Rehovot 76000, Israel
\\
\bf{2} Janelia Research Campus, Howard Hughes Medical Institute, Ashburn, Virginia 20147
\\
\bf{3} The Simons Center for Systems Biology, Institute for Advanced Study, Princeton, NJ 08540
\\
\bf{+} these authors contributed equally to this work
\\
\bigskip
* misha@weizmann.ac.il

\end{flushleft}

\vspace{1cm}

\section*{Abstract}

Human memory appears to be fragile and unpredictable. Free recall of random lists of words is a standard paradigm used to probe episodic memory. We proposed an associative search process that can be reduced to a deterministic walk on random graphs defined by the structure of memory representations. The corresponding graph model can be solved analytically, resulting in a novel parameter-free prediction for the average number of memory items recalled ($R$) out of $M$ items in memory: $R = \sqrt{3\pi M/2}$. This prediction was verified with a specially designed experimental protocol combining large-scale crowd-sourced free recall and recognition experiments with randomly assembled lists of words or common facts. Our results show that human memory can be described by universal laws derived from first principles. 

\vspace{0.25cm}
Keywords: Model, Neural Network, Free recall, Working Memory, Theory
\vspace{0.25cm}



Human cognition is typically considered to be too complex to be described by physics-style universal mathematical laws (see a notable exception in the form of a universal law of generalization proposed in \citealt{shepard87}). Human memory in particular is a critically important mental capacity that includes multiple processes, most crucially acquisition, maintenance and recall (see e.g. \citealt{dudai2004}). While human memory capacity for information is practically infinite, all of the mentioned processes are not entirely reliable, for example recall is often a challenging task even when information being recalled is encoded in memory. An important advantage for studying recall is that it can be precisely quantified with a classical paradigm of 'free recall' (see e.g. \citealt{kahana2012foundations}). Typical experiments involve recalling randomly assembled lists of words in an arbitrary order after a brief exposure. When the presented list becomes longer, the average number of recalled words grows but in a sublinear way (\citealt{binet1894memoire,standing1973learning,murray1976standing}). The exact mathematical form of this relation is controversial and was found to depend on the details of experimental procedures, such as presentation rate (\citealt{waugh1967presentation}). In some studies, recall performance was found to exhibit a power-law relation to the number of presented words (\citealt{murray1976standing}), but parameters of this relation were extremely variable across different experimental conditions. 

These observations seem to rule out any possibility that memory recall can be described by a universal mathematical law that would hold for all experimental conditions and all people. Yet in this study we demonstrate with new experiments that most of the variability in recall can be accounted for by measuring the acquisition and maintenance of information during the presentation phase of the experiment, and when that is controlled, recall itself is much more predictable. Moreover, relation between the number of items in memory and the average fraction of it that can be successfully recalled is described by a parameter-free analytical expression derived here from the deterministic model introduced in (\citealt{romani2013scaling,katkov2017memory}). In other words, despite the overall unpredictability of human memory, some aspects of it obey simple universal laws.

\begin{wrapfigure}[22]{l}{75mm}
\subfloat{\label{fig:graphModela}}
\subfloat{\label{fig:graphModelb}}
\subfloat{\label{fig:graphModelc}}
\centering
\vspace{-.7cm}
\includegraphics[width=\linewidth]{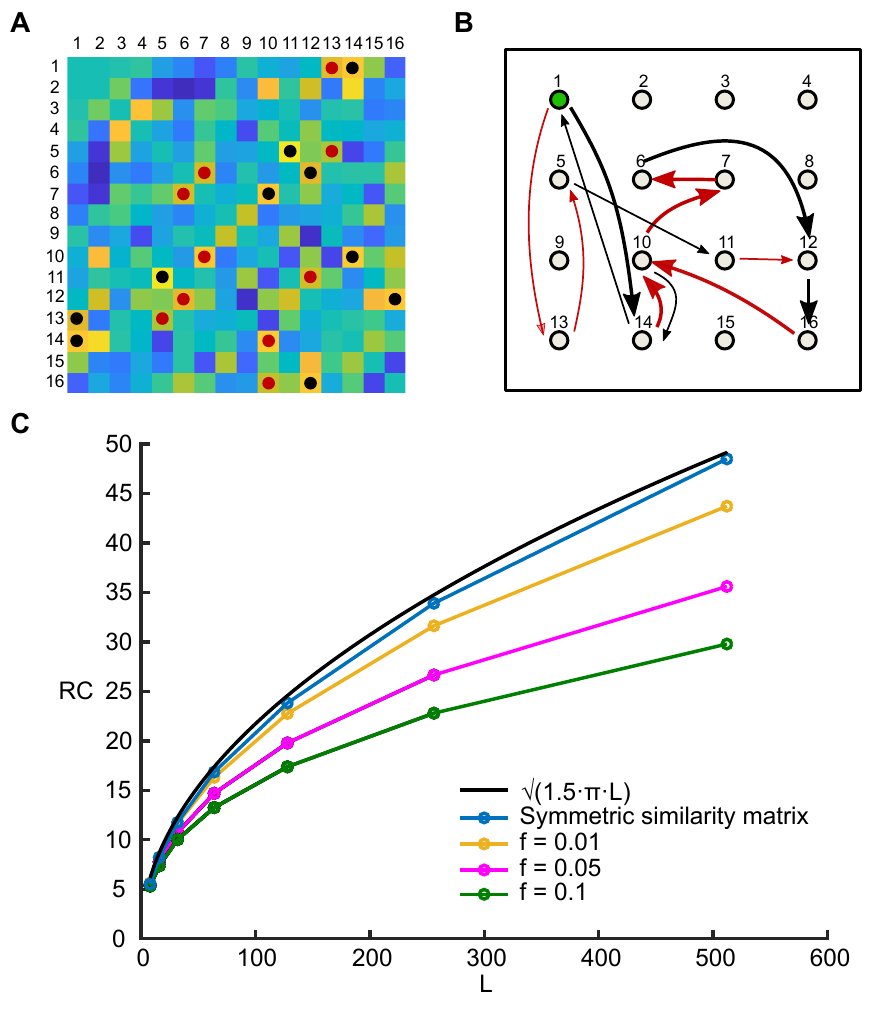}
\captionsetup{labelformat=empty} 
\caption{} 
\label{fig:graphModel}
\end{wrapfigure}
\marginpar{
\vspace{0.3cm} 
\color{Gray} 
\textbf{Figure \ref{fig:graphModel}. Associative search model of free recall. } 
\\(\textbf{A}) SM (similarity matrix) for a list of $16$ items (schematic). For each recalled item, the maximal element in the corresponding row is marked with a black spot, while the second maximal element is marked with a red spot. \\(\textbf{B}) A graph with $16$ nodes illustrates the items in the list. Recall trajectory begins with the first node, and proceeds to an item with the largest similarity to the current one (black arrow) or the second largest one (red arrow) if the item with the largest similarity is the one recalled just before the current one. When the process returns to the 10th item, a second sub-trajectory is opened up (shown with thinner arrows) and converges to a cycle after reaching the $12^\mathrm{th}$ node for the second time. \\(\textbf{C}) Comparison between simulations with random symmetric similarity matrix (blue line) and SM defined by random sparse ensembles with sparsity $f=0.01$ (yellow line), $f=0.05$ (magenta line), $f=0.1$ (green line) and $N = 100000$ number of neurons. Each point is the mean of $10000$ simulations. Black line corresponds to theoretical $ \sqrt{\frac{3}{2} \pi L}$.
}

The proposed recall process is based on two principles:

\begin{itemize}
    \item Memory items are represented in the brain by overlapping random sparse neuronal ensembles in dedicated memory networks;
    \item The next item to be recalled is the one with a largest overlap to the current one, excluding the item that was recalled on the previous step.
\end{itemize} 

\noindent The~first~principle is~a~common~element of most neural network models of memory (see e.g. \cite{hopfield1982neural,tsodyks1988enhanced}), while the second one is inspired by ``Search of Associative Memory'' (SAM, elaborated later). More specifically, item representations are chosen as random binary \{0,1\} vectors where each element of the vector chosen to be 1 with small probability $f \ll 1$ independently of other elements. Overlaps are defined as scalar products between these representations. The model is illustrated in Fig.~\ref{fig:graphModel} (more details in Supplemental Material), where the matrix of overlaps (`similarity matrix', or SM) between $16$ memory representations is shown in Fig. ~\ref{fig:graphModela}. Fig.~\ref{fig:graphModelb} is a graph that shows the transitions between memory items induced by the SM. When the first item is recalled (say the $1$st one in the list), the corresponding row of the matrix, which includes the overlaps of this item with all the others, is searched for the maximal element ($ 14^{\mbox{th}} $ element in this case), and hence the $ 14^{\mbox{th}} $ item is recalled next. This process continues according to the above rule (black arrows), unless it points to an item that was just recalled in the previous step, in which case the next largest overlap is searched (red arrows). After a certain number of transitions, this process begins to cycle over already visited items. This happens either the first time a previously recalled item is reached again, or the process could make some number of transitions over previously recalled items (items $10,14,1$ in Fig.~\ref{fig:graphModelb}) to open up a new trajectory (items $13,5,11,12$) until finally converging to a cycle. After the cycle is reached, no new items can be recalled. 

In our previous publication (\citealt{romani2013scaling}) we showed that the average number of recalled items (recall capacity, or $R$) scales as a power-law function of the number of items in the list, $L$ with exponent that depends on sparseness parameter $f$. Here we focus on the sparse limit of this model, $f \ll 1$, when one can neglect the correlations between different elements of the SM and replace it by a \textit{random symmetric} matrix (see e.g. \cite{quian2010measuring}, for biological motivation for considering a very sparse encoding). We show below that while the corresponding graph model has a history-dependent transition rule and hence is more complex than the standard family of graphs resulting from random mappings (see e.g.\cite{harris1960probability}), it can still be solved analytically in terms of the average number of items visited before converging to a cycle.   

It is instructive to first consider the simpler case of a fully random asymmetric SM with independent elements. In this case, transitions between any two items are equally likely, with probability $1/(L-1)$. When an item is reached for the second time the process enters into a cycle. Therefore the probability that $k$ out of $L$ items will be retrieved is simply

\begin{equation}
\begin{split}
P\left(k; L\right) &= \left(1 - \frac{1}{L-1}\right)\left(1 - \frac{2}{L-1}\right)...\left(1-\frac{k-2}{L-1}\right)\frac{k-1}{L-1} \\
& \simeq \frac{k}{L}e^{- \sum_{i=1}^{k} \frac{k}{L}} \simeq \frac{k}{L} e^{- \frac{k^2}{2L}}
\end{split}
\end{equation}

\noindent where we considered a limit of large number of items in the list ($L \gg 1$) and assumed that $L \gg k \gg 1$, which is confirmed a posteriori below. The average number of recalled words can then be calculated as 

\begin{equation}
\begin{split}
    R &= \langle k \rangle = \sum_{k=2}^L  k  \frac{k}{L} e^{- \frac{k^2}{2L}} \\
    & \approx  \sqrt{L} \int_0^\infty x^2 e^{- \frac{x^2}{2}} dx \\
    &= \sqrt{\frac{\pi L}{2}} 
\end{split}
\label{eq:random}
\end{equation}

\noindent which is a well known result in random graphs literature (\cite{harris1960probability,katz1996handbook}).

When the SM is symmetric, as in our case, the statistics of transitions in the corresponding graph is more complicated (see Supplemental Material for more details about the derivation). In particular, the probability for a transition to one of the previously recalled items scales as $1/(2L)$ rather than $1/L$ as in the case of asymmetric SMs, and hence the average length of trajectory until the first return converges to $\sqrt{\pi L}$. Moreover, with probability $1/3$ the trajectory then turns towards previous items and opens up a new route until again hitting a previously recalled item, etc. Taken together, the chance that recall trajectory enters a cycle after each step asymptotically equals to $1/(2L) \cdot 2/3 = 1/(3L)$, as opposed to $1/L$ for the fully random matrix, and hence the $R$ can be obtained by replacing $L$ by $3L$ in Eq.~\eqref{eq:random}:

\begin{equation}
    R \approx \sqrt{\frac {3\pi}{2} L} \approx 2.17 \cdot \sqrt{L} \, ,
\label{eq:Romani}
\end{equation}

\noindent see Fig.~\ref{fig:graphModelc} for the comparison of this analytical estimate with numerical simulations of the model. We emphasize that Eq.~\eqref{eq:Romani} does not have any free parameters that could be tuned to fit the experimental results. Hence, both the exponent and coefficient of this power law expression are a result of the assumed recall mechanism; in other words, this equation constitutes a true prediction regarding the asymptotic recall performance for long lists of items as opposed to earlier theoretical studies. Here we present the results of our experiments designed to test this prediction. 

$ $

\begin{wrapfigure}[23]{l}{75mm}
\subfloat{\label{fig:experimenta}}
\subfloat{\label{fig:experimentb}}
\subfloat{\label{fig:experimentc}}
\centering
\vspace{-.8cm}
\includegraphics[width=\linewidth]{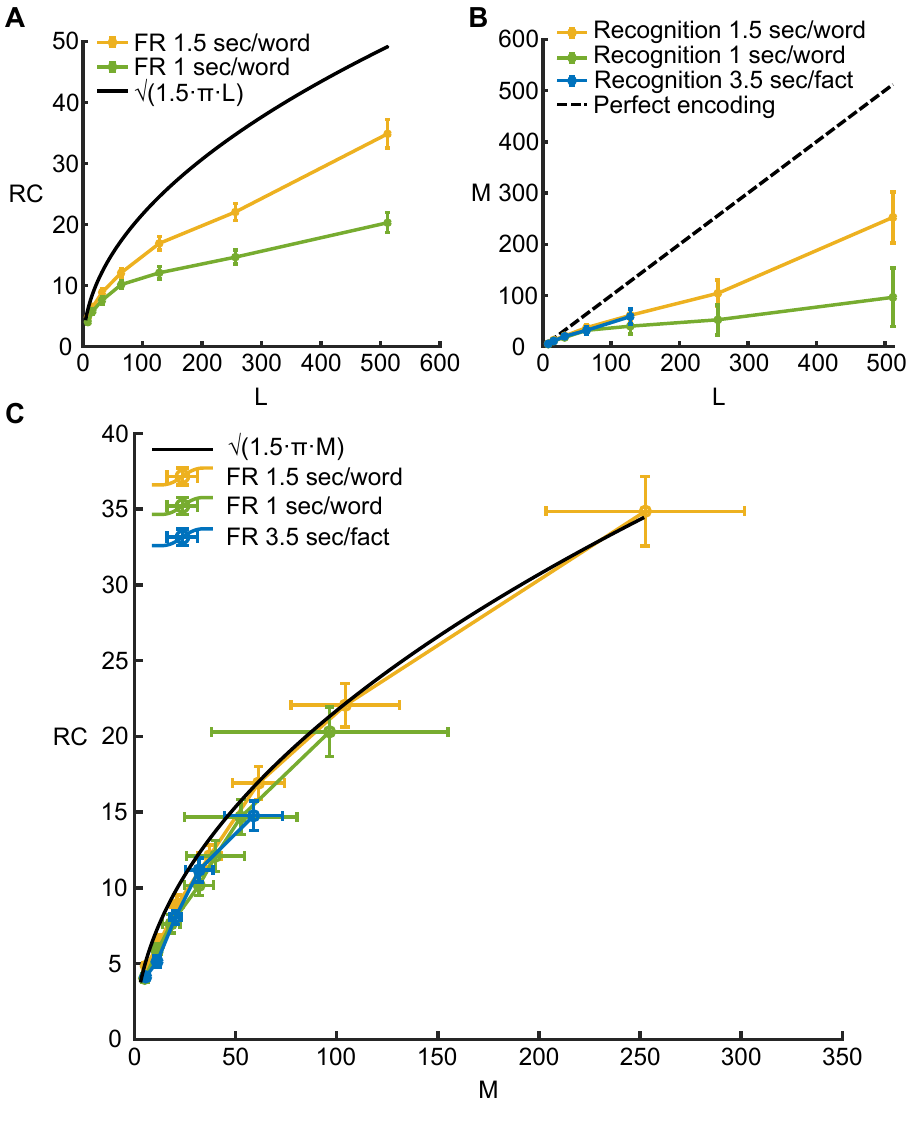}
\captionsetup{labelformat=empty} 
\caption{} 
\label{fig:experiment}
\end{wrapfigure}
\marginpar{
\vspace{.2cm} 
\color{Gray} 
\textbf{Figure \ref{fig:experiment}. Human recall and recognition performance. } 
\\ (\textbf{A}) Average number of words recalled as a function of the number of words presented. Black line: Eq.~\eqref{eq:Romani}. Yellow line: experimental results for presentation rate $1.5$ sec/word. Green line: experimental results for presentation rate $1$ sec/word. The error in $R$ is a standard error of the mean. \\ (\textbf{B}) Estimated average number of encoded words/sentences for lists of different lengths. Black dashed line corresponds to perfect encoding, green line corresponds to presentation rate $1$ sec/word and yellow line to presentation rate $1.5$ sec/word; blue line corresponds to lists of short sentences (see text for details). The error in $M$ is computed with bootstrap procedure (\citealt{efron1994introduction}). \\ (\textbf{C}) Average number of words/sentences recalled as a function of the average number of encoded words. Black line: theoretical prediction, Eq.~\eqref{eq:Naim}. Green line: experimental results for presentation rate $1$ sec/word. Yellow line: experimental results for presentation rate $1.5$ sec/word. Blue line: experimental results for short sentences. The error in $R$ is a standard error of the mean, while the error in $M$ is computed with bootstrap procedure (see Supplemental Material for details).
}

The universality of the above analytical expression for $R$ seems to contradict our everyday observations that people differ in terms of their memory effectiveness depending, e.g. on their age and experience. Moreover, it is at odds with previous experimental studies showing that performance in free recall task strongly depends on the experimental protocol, for example presentation rate during the acquisition stage (see e.g. \citealt{murdock1960immediate,murdock1962serial,roberts1972free,howard1999contextual,kahana2002age,zaromb2006temporal,ward2010examining,miller2012recall,grenfell2017common}) and the extent of practice (\citealt{klein2005comparative,romani2016practice}). Since most of the published studies only considered a limited range of list lengths, we performed free recall experiments on the Amazon Mechanical Turk\textsuperscript{\textregistered} platform for list lengths of $8, 16, 32, 64, 128, 256$ and $512$ words, and two presentation rates: $1$ and $1.5$ seconds per word. To avoid practice effects, each participant performed a single free recall trial with a randomly assembled list of words of a given length. The results confirm previous observations that recall performance improves as the time allotted for acquisition of each word increases, approaching the theoretical prediction of Eq.~\eqref{eq:Romani} from below (see Fig.~\ref{fig:experimenta}). 

We reasoned that some or all of the variability in the experimentally observed $R$ could result from the variability in the number of words that remain in memory as candidates for recall after the list is presented. In particular, some of the words could be missed at presentation, while others could be acquired but later erased or degraded. It seems reasonable that acquisition depends on various factors, such as attention, age of participants, acquisition speed, etc. One should then correct Eq.~\eqref{eq:Romani} for $R$, replacing the number of presented words $L$ with the number of words in memory after the whole list is presented, $M$:

\begin{equation}
\begin{split}
    R \approx \sqrt{\frac{3 \pi}{2}M} 
\end{split}
\label{eq:Naim}
\end{equation}

\noindent To~test~this~conjecture,~we~designed a novel experimental protocol that involved performing both recall and {\it recognition} experiments on the same group of participants. Each participant performed one recognition and one recall trial with lists of the same number of words (but different words between recognition and recall) and under identical presentation conditions, including presentation rate, in order to independently evaluate the average number of words in memory, and the average number of words recalled. Following \cite{standing1973learning}, at the end of presentation we showed each participant a pair of words, one from the list just presented (target) and one randomly chosen lure, requesting to report which word was from the list. The average number of words remaining in memory ($M$) was then estimated from the fraction of correctly recognized words ($c$) by assuming that if a target word was still encoded at the end of presentation, it will be chosen during recognition test, otherwise the participant will randomly guess which of the two words is a target: $M=L \cdot (2c-1)$. Importantly, each participant performed a single recognition test, to avoid the well known effect of `output interference' between subsequent recognition tests for a single list (see e.g. \citealt{criss2011output}).

Fig.~\ref{fig:experimentb} shows the estimated average $M$ as a function of list length $L$ (see Supplemental Material for details of analysis). Results confirm that acquisition improves with time allotted to presentation of each word. Standard error of the mean for the number of encoded words across participants, for each list length and each presentation speed, was estimated with a bootstrap procedure by randomly sampling a list of participants with replacement (\citealt{efron1994introduction}, see Supplemental Material). 

\noindent In Fig.~\ref{fig:experimentc} experimentally obtained $R$ (yellow and green lines) is compared with the theoretical prediction of Eq.~\eqref{eq:Naim} (black line), where $M$ is the average number of encoded words, estimated in the recognition experiment. Remarkably, agreement between the data and theoretical prediction is very good for both presentation rates, even though the number of encoded and recalled words is very different in these two conditions for each value of list length. We also performed multiple simulations of our recall algorithm 
(\citealt{romani2013scaling,katkov2017memory}) and found that it captures the statistics of the recall performances as accessed with bootstrap analysis of the results (see Fig.~\ref{fig:simulation_and_bootstrap} in Supplemental Material). 

Experiments presented above, as well as the vast majority of previous recall experiments, were performed with lists of words. To test the generality of our model prediction, we generated a set of $325$ short sentences expressing common knowledge facts, such as `Earth is round' or `Italians eat pizza', etc. We repeated our experiments with random lists of $8, 16, 32, 64,$ and $128$ such sentences, each presented for $3.5$ seconds (see Supplementary Material for more details of the analysis). As shown in Figs.~\ref{fig:experimentb} and \ref{fig:experimentc}, performance with lists of sentences is very close to that of words with $1.5$ words per second presentation rate, albeit with some small deviations towards lower levels.

\section*{Discussion}

The results presented in this study show that the relation between the number of words in memory and the number of recalled words conforms with remarkable precision to the analytical, parameter-free expression Eq.~\eqref{eq:Naim}, derived from a deterministic associative search model of recall. The relation between these two independently measured quantities holds even though both of them strongly depend on the number of presented words and on the presentation rate. We further confirmed the generality of Eq. \eqref{eq:Naim} by repeating the experiments with lists of short sentences expressing common knowledge facts. Hence it appears that memory recall is a more universal process than memory acquisition and maintenance. The crucial aspect of the model is the similarity matrix between the items that determines the recall transitions, but the precise nature of this matrix beyond its statistics across the presented lists and/or across participants does not have to be specified. It seems plausible that different people will have different similarity matrices, reflecting their unique language experience, which makes direct estimation of it rather challenging. However, our previous study (\cite{recanatesi2015neural}, see also \cite{howard2007semantic}) showed that recall transitions are sensitive to the measure of semantic similarity called Latent Semantic Analysis (LSA), which represents the number of times two words appear together in a representative corpora of natural text (\cite{landauer1997solution}). This indicates that there is some degree of universality in inter-word similarities across all people.  

Several influential computational models of recall were developed in cognitive psychology that incorporate interactive probabilistic search processes (see e.g. \citealt{raaijmakers1980sam,gillund1984retrieval,howard2002distributed,laming2009failure,polyn2009context,lehman2013buffer}). These cognitive models have multiple free parameters that can be tuned to reproduce the experimental results on recall quite precisely, including not only the number of words recalled but also the temporal regularities of recall, such as primacy, recency and temporal contiguity effects (\citealt{murdock1962serial,murdock1970interresponse,howard1999contextual}). However, most of the free parameters lack clear biological meaning and cannot be constrained before the data is collected, hence the models cannot be used to predict the recall performance but only explain it a posteriori. Our recall model can be viewed as a radically simplified version of the classical `Search of Associative Memory' model (SAM), see \citealt{raaijmakers1980sam}. In both models, recall is triggered by a matrix of associations between the items, which in SAM is built up during presentation according to a rather complex set of processes, while in our model is simply assumed to be a fixed, structure-less symmetric matrix (see Fig.~\ref{fig:graphModel}). Subsequent recall in SAM proceeds as a series of attempted probabilistic sampling and retrievals of memory items, until a certain limiting number of failed attempts is reached after which recall terminates. In our model, this is replaced by a deterministic transition rule that selects the next item with the strongest association to the currently recalled one. As a result, recall of new items terminates automatically when the algorithm begins to cycle over already recalled items, without a need to any arbitrary stopping rule. Finally, SAM assumes that all the presented words are stored into long-term memory to different degrees, i.e. could in principle be recalled, while in the current study we assume that only a certain fraction of words remain in memory at the end of presentation to become candidates for recall. This assumption is confirmed a posteriori by the collapse of $R$ vs $M$ curves for different presentation rates (see Supplementary Material for more detailed argumentation). We also neglected the well-documented effects of short-term memory on free recall (see e.g. \cite{glanzer2mechanisms1966}), which are very small in our data (see Supplementary Material). 

We consider it little short of a mystery that with these radical simplifications, the model predicts the recall performance with such a remarkable precision and without the need to tune a single parameter. This suggests that despite all the simplifications, the model faithfully captures a key first-order effect in the data. Future theoretical and experimental studies should be pursued to probe which aspects of the model are valid and which are crucial for the obtained results. 

\section*{Acknowledgments}

This research has received funding from the European Union’s Horizon 2020 Framework Programme for Research and Innovation under the Specific Grant Agreement No. 785907 (Human Brain Project SGA2); EU-M-GATE 765549 and Foundation Adelis. S.R. is supported by the Howard Hughes Medical Institute. We thank Drs. Mike Kahana and Eli Nelken for helpful comments.

\nolinenumbers

\bibliography{sample}

\bibliographystyle{plainnat}

\clearpage

\clearpage

\end{document}


\manuallabel{eq:random}{2}
\manuallabel{fig:graphModela}{1a}
\manuallabel{fig:graphModelb}{1b}
\manuallabel{fig:graphModelc}{1c}
\manuallabel{fig:graphModel}{1}
\manuallabel{fig:experimenta}{2a}
\manuallabel{fig:experimentb}{2b}
\manuallabel{fig:experimentc}{2c}
\manuallabel{fig:experiment}{2}
\begin{flushleft}
{\Large
\textbf{Supplemental material}
}
\newline

\end{flushleft}

\renewcommand{\theequation}{S\arabic{equation}}

\begin{adjustwidth}{-2in}{0in}
\section*{Methods}

\subsection*{Recall model}

Our recall model is presented in more details in \citealt{romani2013scaling,katkov2017memory}. In this contribution we considered a simplified version of the model, where we approximate the matrix of overlaps between random sparse memory representations by a random symmetric $L$ by $L$  similarity matrix (SM) with otherwise independently distributed elements, where $L$ is a number of words in the list. Neglecting the correlations between SM elements is justified in the limit of very sparse encoding of memory items (see \citealt{romani2013scaling}). A new matrix is constructed for each recall trial. The sequence $\{k_1, k_2, \dots, k_r \}$ of recalled items is defined as follows. Item $k_1$ is chosen randomly among all $L$ presented items with equal probability. When $n$ items are recalled, the next recalled item $k_{n+1}$ is the one that has the maximal similarity with the currently recalled item $k_n$, excluding the item that was recalled just before the current one, $k_{n-1}$. After the same transition between two items is experienced for the second time, the recall is terminated since the model enters into a cycle. 

\subsection*{Solution of the recall model}

The symmetry of SM appears to be a minor difference from the much simpler model of fully random asymmetric SM presented in the main text, but in fact it significantly impacts the statistics of the transitions in the corresponding graphs as we will show below.

If retrieval always proceeds from an item to its most similar, as in the asymmetric case, the dynamics will quickly converge to a two-items loop. The reason is that if item $B$ is most similar to item $A$, then item $A$ will be most similar to item $B$ with a probability of approximately $0.5$. We hence let the system choose the second most similar item if the most similar one has just been retrieved, as explained in the main text. When reaching an already visited item, retrieval can either repeat the original trajectory (resulting in a loop) or continue backward along the already visited items and then open a new sub-trajectory (see Fig.~\ref{fig:graphModelb}). Here we show how to calculate the probability of returning from a new item to any one of already visited items and the probability that the retrieval proceeds along the previous trajectory in the opposite direction upon the return.

In order to return back from item $k$ to item $n$, the $n^{th}$ element of the $k^{th}$ row of SM, $S_{kn}$, has to be the largest of the remaining $L-2$ elements in the $k^{th}$ row (excluding the diagonal and the element corresponding to the item visited just before the $k^{th}$ one). The probability for this would be $\approx \frac{1}{L}$ for an asymmetric matrix. For a symmetric matrix ($S_{nk} = S_{kn}$), we have an additional constraint that the element $S_{kn}$ is \textit{not} the largest in the $n^{th}$ row of $S$, since we require that the $k^{th}$ item was \textit{not} retrieved after the first retrieval of the $n^{th}$ one. The probability of return is then equal to

\begin{equation}
P \left( S_{kn} = max \left( \vec{S_k} \right) | S_{kn} < max \left( \vec{S_n} \right) \right) \approx \frac{1}{2L}
\end{equation}

\noindent where $\vec{S_k}$ denotes the vector of relevant elements in the $k^{th}$ row of matrix $S$. The return probability is therefore reduced by a factor of two due to the symmetric nature of SM but retains the same scaling with $L$ as in the model with asymmetric SM. After the first return to an item $n$ ($=10$ in Fig. \ref{fig:graphModelb} of the main paper), the trajectory may either begin to cycle, or turn towards previously visited items but in the opposite direction if the original transition from this item ($10 \rightarrow 7$ in Fig.~\ref{fig:graphModelb}) was along the second largest element of $\vec{S_n}$. The marginal probability for this is $\frac{1}{2}$, but we must impose the constraint that the $k^{th}$ item was \textit{not} retrieved after the first retrieval of the $n^{th}$ one. If the item preceding $n$ is $j$ ($14$ in Fig. \ref{fig:graphModelb}), the corresponding probability is given by 

\begin{equation}
 P \left( max \left( \vec{S_n} \right) < max \left( \vec{S_j} \right) | max \left( \vec{S_n} \right) > max \left( \vec{S_k} \right) \right) \approx \frac{1}{3} \, ,
\end{equation}

\noindent which follows from the observation that any ordering for the maximal elements of three vectors of equal size is equally probable. From this result, we conclude that the average number of sub-trajectories during the retrieval process is $\frac{3}{2}$. All together the chance for the process to enter a cycle after each new item retrieved is $\frac{1}{2L}\frac{2}{3}=\frac{1}{3L}$ and hence the average number of items recalled is estimated by replacing $L$ with $3L$ in the corresponding expression for RC in the model with fully random asymmetric SM, Eq.~\eqref{eq:random} of the main text:

\begin{equation}
\begin{split}
    & RC = k \cdot \sqrt{L} \\
    & k \approx \sqrt{3\pi/2} \approx 2.17
\end{split}
\end{equation}

\subsection*{Participants, Stimuli and Procedure}

Experiments and data analysis were performed at the Weizmann Institute of Science. Ethics approval was obtained by the IRB (Institutional Review Board) of the Weizmann Institute of Science. Each participant accepted an informed consent form before participation and was receiving from $50$ to $85$ cents for approximately $5-25$ min, depending on the task. In total $1039$ participants, were recruited to perform memory experiments on the Amazon Mechanical Turk\textsuperscript{\textregistered} (https://www.mturk.com). All participants were previously selected in the lab Prof. Mike Kahana from the University of Pennsylvania (private communication). Presented lists were composed of non-repeating words randomly selected from a pool of $751$ words produced by selecting English words (\citealt{healey2014individual}) and then maintaining only the words with a frequency per million greater than $10$ (\citealt{medler2005mcword}). The stimuli were presented on the standard Amazon Mechanical Turk\textsuperscript{\textregistered} web page for Human Intelligent Task. Each trial was initiated by the participant by pressing ``Start Experiment'' button on computer screen. List presentation followed $300$ ms of white frame. Each word was shown within a frame with black font for $500$ or $1000$ ms (depending on presentation rate) followed by empty frame for $500$ ms. After the last word in the list, there was a $1000$ ms delay before participant performed the task. The set of list lengths was: $8$, $16$, $32$, $64$, $128$, $256$ and $512$ words. 
We also performed the same experiments using a set of $325$ short sentences expressing well-know facts, such as `Earth is round' or `Italians eat pizza', etc.  We repeated our experiments with random lists of $8, 16, 32, 64,$ and $128$ such sentences, each presented for $2500$ ms followed by empty frame for $1000$ ms. Each participant performed experiment A (free recall) and Experiment B (recognition) with lists of the same length. In more details

\begin{itemize}
    \item $348$ participants performed the two experiments with presentation rate of $1.5$ sec/word: $265$ participants did both experiments for only one list length, $54$ for two list lengths, $18$, $9$ and $2$ for $3$, $4$ and $5$ list lengths respectively. 
    \item $375$ participants performed the two experiments with presentation rate of $1$ sec/word: $373$ participants did both experiments for only one list length, $2$ for two list lengths.
    \item $331$ participants performed the two experiments with presentation rate of $3.5$ sec/fact: $328$ participants did both experiments for only one list length, $3$ for two list lengths. $15$ participants performed also experiments with the words.
\end{itemize}

\noindent \textbf{Experiment A - Free recall}.
Participants were instructed to attend closely to the stimuli in preparation for the recalling memory test. After presentation and after clicking a ``Start Recall'' button, participants were requested to type in as many items (words/sentences) as they could in any order. After the finishing the typing (following non-character input) the information was erased from the screen, such that participants were seeing only the currently typed item. Only one trial was performed by each participant. The time for recalling depended on the length of the learning set, from $1$ minute and $30$ seconds up to $10$ minute and $30$ seconds, with a $1$ minute and $30$ seconds increase for every length doubling. The obvious misspelling errors were corrected. Repetitions and the intrusions (items that were not in the presented list) were ignored during analysis.

\medbreak

\noindent \textbf{Experiment B - Recognition task}.
In recognition trial, after presentation and after clicking a ``Start Recognition'' button, participants were shown $2$ items, one on top of another. One item was randomly selected among just presented in the list (target), and another one was selected from the rest of the pool of words or sentences. The vertical placement of the target was random.  Participants were requested to click on the items they think was presented to them during the trial. Each list was followed with $5$ recognition trials per participant, but only the first trial was considered in the analysis. Time for all trials was limited to $45$ min, but in practice each response usually took less than two seconds.

\subsection*{Analysis of the results}

The average number of recalled items ($RC$) for each list length and its standard error were obtained from the distribution of the number of recalled items across participants. 

In the case of sentences, additional problem that arose concerned the different possible phrasing of the same facts. For example, if a fact presented was `Italians like pizza' and a participant reported `Pizza is loved by Italians', we had to find a way to identify it as correctly recalled. To this end, we used word2vec software developed by Google (\citealt{mikolov2013distributed}). Word2vec is a group of related models that are used to produce multidimensional word embeddings. These models are shallow, two-layer neural networks that are trained to reconstruct linguistic contexts of words. Word2vec takes as its input a large corpus of text and produces a vector space, typically of several hundred dimensions, with each unique word in the corpus being assigned a corresponding vector in the space. Word vectors are positioned in the vector space such that words that share common contexts in the corpus are located in close proximity to one another in the space. We used word2vec to compare the sentences reported by participants to the ones presented. A sentence vector was computed as the average of all the word vectors in the sentence, and the similarity between any two sentences with vectors $S_1$ and $S_2$ was defined via the cosine of the angle between them, as $1 - \cos{(S_1,S_2)}$. If the similarity was greater than $0.9$ the recall was considered to be correct (this threshold was confirmed by manual inspections of multiple cases). The cases in which the similarity was between $0.8$ and $0.9$ were checked manually.

The average number of items that remain in memory after presentation of a list of length $L$ was computed from the results of recognition experiments as in (\citealt{standing1973learning}). Suppose that $M$ out of $L$ items are remembered on average after an exposure to the list, the rest are lost. The chance that one of the remembered items is presented during a recognition trial is then $M/L$, while the chance that a lost word is presented is $1-M/L$. We assume that in the first case, a participant correctly points to a target item, while in the second case, she/he is guessing. The fraction of correct responses $c$ can then be computed as

\begin{equation}
c = \frac{M}{L} + \frac{1}{2} \cdot \left( 1 - \frac{M}{L} \right) \, .
\end{equation}

\noindent Hence the average number of remembered items can be computed as

\begin{equation}
M = L \cdot \left( 2 c - 1 \right) \, .
\label{eq:EqStandind}
\end{equation}

The above analysis is based on a simplified assumption that after the list is presented, each word either remains in memory or completely erased. Our experiments cannot shed light on what really happens to words that are not recognized as familiar after the presentation of the list – they could have never been encoded in memory in the first place, encoded but then erased completely or alternatively degraded to the degree that prevents recognition but not fully erased. Different psychological studies accept different assumptions, e.g. Standing 1973 paper talks about ‘words that remain in memory’, indicating all-or-none processes, while SAM model assumes that all presented words remain in memory but at different degrees. In order to test our conjecture that dependence of recall on presentation speed can be explained away by the different number of words that are candidates for recall, we assumed that the latter can be reliably estimated by our recognition experiments, i.e. that words that could not be recognized could also not be recalled, which may or may not necessarily imply all-or-none encoding/forgetting. We believe that our experimental results a posteriori support this conjecture. In particular, if our estimates for the number of words that remain recall candidates after the whole list is presented (M) was not good, or meaningless, there would be no reason for our RC vs M data to collapse to practically the same curve for two presentation rates, given that both RC and M are dramatically different for these two conditions. 

In order to estimate a standard error of the mean for the number of remembered items across participants, for each list length, we performed a bootstrap procedure (\citealt{efron1994introduction}). We generated multiple bootstrap samples by randomly sampling a list of N participants with replacement N times. Each bootstrap sample differs from the original list in that some participants are included several times while others are missing. For each bootstrap sample $b$ out of total number $B$, with $B=500$, we compute the estimate for the average number of remembered items, $M(b)$, according to Eq.~\eqref{eq:EqStandind}. The standard error of $M$ is then calculated as a sample standard deviation of $B$ values of $M(b)$:

\begin{equation}
  se_B = \sqrt{\sum_{b=1}^B \frac{\left(M \left( b \right) - \bar{M} \right)^2}{B-1}} \, ,
\end{equation}

\noindent where $\bar{M} = \sum_{b=1}^B \frac{M \left( b \right)}{B}$.

\subsection*{Effects of short-term memory on free recall}

It is well known from the previous literature that short-term memory (STM) plays a certain role in free recall, as in particular reflected in pronounced recency effects, i.e. increased probability to recall one of the words in the end of a list that are observed if recall begins immediately after presentation but not if it is delayed (see e.g. \cite{glanzer2mechanisms1966}). To evaluate the influence of STM in our data, we computed the serial position curves (\cite{murdock1962serial}) and observed increased probability of recall at the end of the list, especially for longer lists, mostly for the last one to three serial positions (Fig. \ref{fig:recency}). The overall effect of STM on recall was quite small for all conditions, less then one extra word recalled.

\section*{Additional figures}
\end{adjustwidth}

\setcounter{figure}{0}
\renewcommand{\thefigure}{S\arabic{figure}}

\setcounter{table}{0} 
\renewcommand{\thetable}{S\arabic{table}}

\begin{adjustwidth}{-2in}{0in}
\begin{flushright}
\includegraphics[width=183mm]{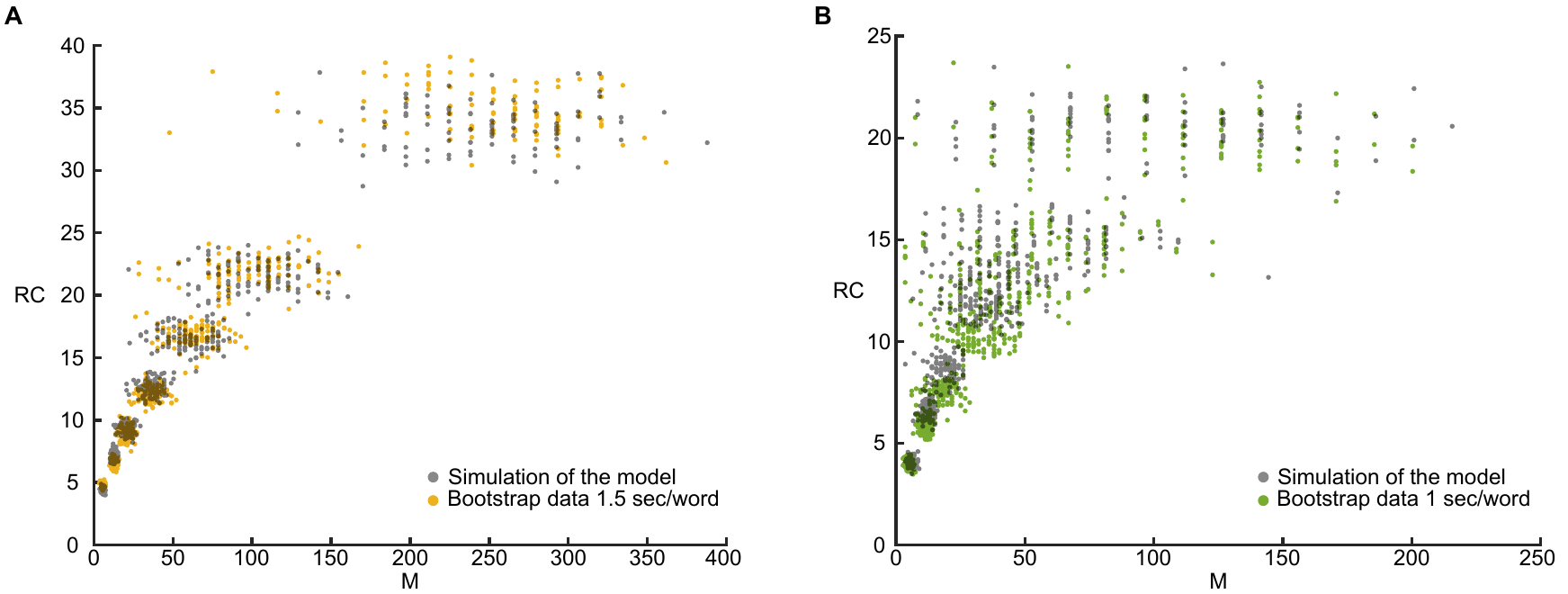}
\end{flushright}
\justify 
\color{Gray}
\textbf {Figure \ref{fig:simulation_and_bootstrap}. Bootstrap analysis and comparison to model simulations.}
\\ \noindent (\textbf{A}) 1.5 seconds per word presentation rate; (\textbf{B}) 1 second per word presentation rate. 
\\  \noindent 100 bootstrap samples for each list length are shown with colored dots with coordinates $M(b)$ and $RC(b)$, where $RC(b)$ is an average number of recalled words computed for each bootstrap sample $b$. Black dots show corresponding simulation results, obtained as follows. From the results of recognition experiment, we calculate, for each list length $L$, the fraction of correct recognitions across the participants, $c$, and therefore the probability $p = \left(2c -1 \right)$ that a presented word is remembered. With these two numbers, we simulate multiple recognition and recall experiments. For recognition experiment, we draw a binomial random variable with probability $c$ for each participant independently, simulating their recognition answers, from which we compute the number of remembered words averaged for all participants as explained in the Methods. We then drew $L$ binomial variables with probability $p$ for each participant, simulating the number of remembered of words by this participant during the recall experiment. With the number of remembered words known for each participant, we run the recall model (see Methods) to obtain the average recall performance over participants. Every simulation described above produced $7$ pairs of results $\left(M,RC\right)$, one per list length. We repeated the whole procedure $100$ times, same as the number of bootstrap samples.
\end{adjustwidth}

\begin{figure}[h]
\subfloat{\label{fig:simulation_and_bootstrapa}}
\subfloat{\label{fig:simulation_and_bootstrapb}}
\centering
\captionsetup{labelformat=empty} 
\caption{} 
\label{fig:simulation_and_bootstrap}
\end{figure}

\begin{adjustwidth}{-2in}{0in}
\begin{flushright}
\includegraphics[width=183mm]{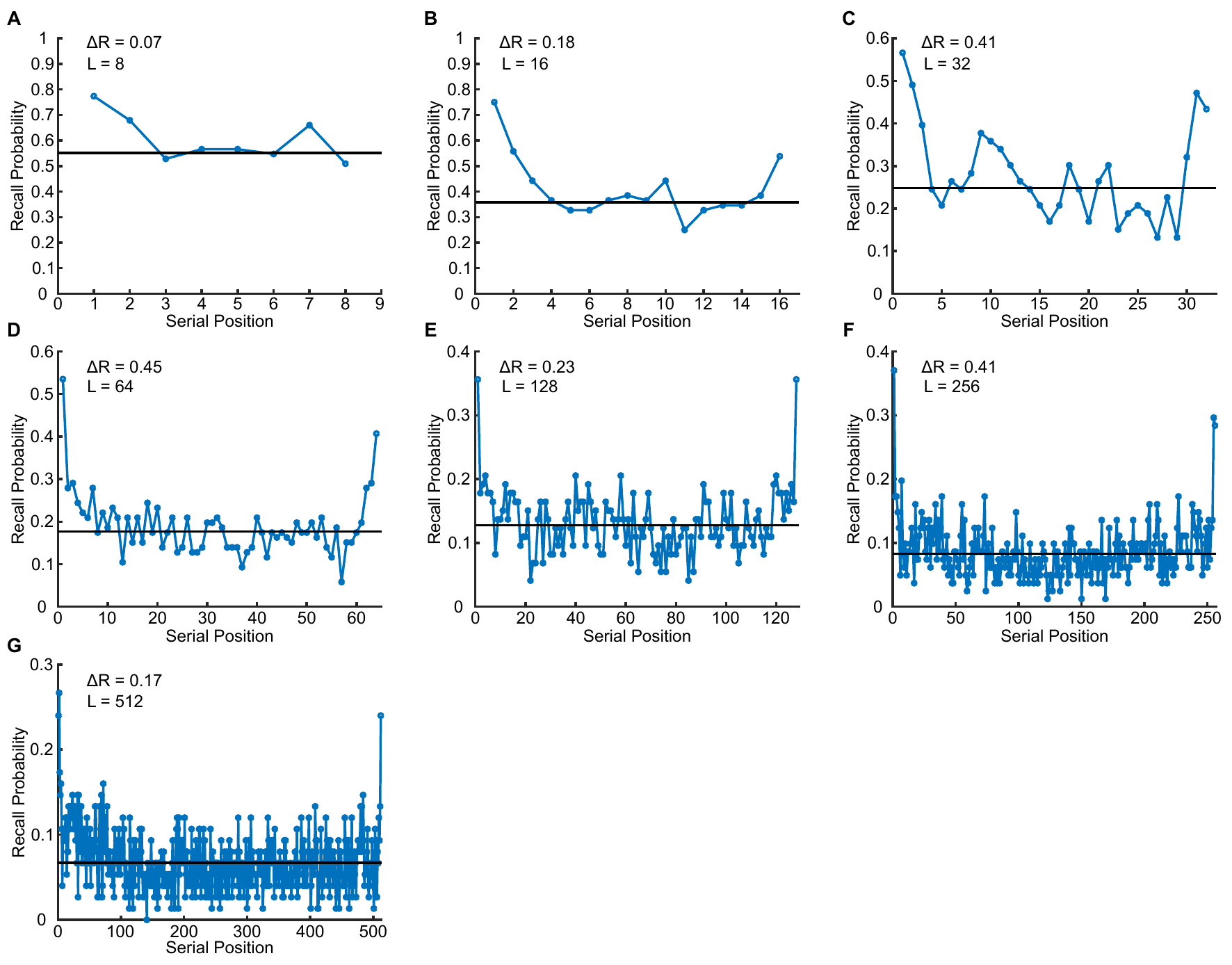}
\end{flushright}
\justify 
\color{Gray}
\textbf {Figure \ref{fig:recency}. Effects of short-term memory on free recall.}
\\ \noindent Probability to recall a word as a function for its serial position in the presented list, for presentation rate of $1.5$ sec per word. Black horizontal curve illustrates the average recall probability that is computing by excluding first $3$ and last $2$ words in the list. The additional number of words recalled ($\Delta RC$) is computed by summing the excess recall probability for $1$ to $3$ words in the end of the list that are significantly better recalled than the rest. 
\end{adjustwidth}

\begin{figure}[h]
\subfloat{\label{fig:reviewa}}
\subfloat{\label{fig:reviewb}}
\subfloat{\label{fig:reviewc}}
\subfloat{\label{fig:reviewd}}
\subfloat{\label{fig:reviewe}}
\subfloat{\label{fig:reviewf}}
\subfloat{\label{fig:reviewg}}
\centering
\captionsetup{labelformat=empty} 
\caption{} 
\label{fig:recency}
\end{figure}

\clearpage

\begin{adjustwidth}{-2in}{0in}

\bibliography{sample}

\bibliographystyle{plainnat}
\end{adjustwidth}